\documentstyle[aps,epsf,manuscript,eqsecnum]{revtex}
\input epsf
\def\sn2t{\sin^22\theta}
\def\tn2t{\tan^22\theta}
\def\UU{Y_U^\dagger Y_U}
\def\DD{Y_D^\dagger Y_D}
\def\EE{Y_E^\dagger Y_E}
\def\dmdt{16\pi^2{dm\over dt}}
\def\mee{m_{ee}^0}
\def\memu{m_{e\mu}^0}
\def\metau{m_{e\tau}^0}
\def\mmue{m_{\mu e}^0}
\def\mmumu{m_{\mu\mu}^0}
\def\mmutau{m_{\mu\tau}^0}
\def\mtaue{m_{\tau e}^0}
\def\mtaumu{m_{\tau\mu}^0}
\def\mtautau{m_{\tau\tau}^0}
\def\mmutaut{m_{\mu\tau}(t)}
\def\mmumut{m_{\mu\mu}(t)}
\def\mtautaut{m_{\tau\tau}(t)}
\def\be{\begin{equation}}
\def\ee{\end{equation}}
\def\ba{\begin{eqnarray}}
\def\ea{\end{eqnarray}}
\def\br{\begin{array}}
\def\er{\end{array}}

\begin{document}

\rightline{IC/2000/160}
\rightline{UMD-PP-01-21}
\rightline{NEHU/PHY-MP-04/2000}

\begin{center}
{\large \bf Large Neutrino Mixing from Renormalization Group Evolution}

\medskip

{K. R. S. Balaji\footnote{balaji@imsc.ernet.in}}

{\it The Abdus Salam ICTP, 11 Strada Costiera, 34014 Trieste, Italy.}

{and}

{\it Institute of Mathematical Sciences, Chennai 600 113, India.}

\smallskip

{R. N. Mohapatra\footnote{rmohapat@physics.umd.edu}}

{ \it Department of Physics, University of Maryland, 
College park MD 20742, USA.} 

\smallskip

{M. K. Parida\footnote{mparida@dte.vsnl.net.in}}

{\it Department of Physics, North Eastern Hill University, Shillong
793022, India.} 

\smallskip
E. A. Paschos\footnote{paschos@hal1.physik.uni-dortmund.de}

{\it Instit\"{u}t f\"{u}r Physik, Universit\"{a}t Dortmund,
D-44211 Dortmund, Germany.}

\date{\today}

\end{center}

\begin{abstract}
The renormalization group evolution equation for two neutrino mixing is known
to exhibit nontrivial fixed point structure corresponding to maximal mixing at
the weak scale. The presence of the fixed point provides a natural 
explanation of the observed maximal mixing of $\nu_{\mu}-\nu_{\tau}$ if 
the $\nu_{\mu}$ and $\nu_{\tau}$ are assumed to be quasi-degenerate at the
seesaw scale without constraining on the mixing angles at that
scale. In particular, it allows them to be similar to the quark mixings as
in generic grand unified theories. We discuss
implementation of this program in the case of MSSM and find that the predicted
mixing remains stable and close to its maximal value, for all energies below 
the $O$(TeV) SUSY scale. We also discuss how a particular realization
of this idea can be tested in
neutrinoless double beta decay experiments.
\end{abstract}
\vspace{2cm}
\leftline{PACS numbers: 12.15.Ff, 12.60.-i} 

\newpage

\section{INTRODUCTION}

\label{sec1}
\par
Theoretical understanding of experimentally measured neutrino anomalies poses
a major challenge to unified gauge theories especially since
$\nu_\mu-\nu_\tau$ mixing has been observed to be close to maximal through
atmospheric neutrino flux measurements whereas the mixing in the corresponding
quark sector is small. The problem is so severe that, only over the limited
span of the last two years, nearly a hundred models have been proposed where
considerable effort has been devoted to accommodate large neutrino mixing
\cite{ref1}. There are also interesting suggestions to understand this 
large mixing in the context of various grand unified theories including
$SO(10)$ \cite{ref2,ref3}, which unify both quarks and leptons. It is
however fair to say that no convincing and widely accepted natural model
has yet emerged.
\par
With a view to simplifying model building, we recently suggested criteria
for radiative magnification
of neutrino mixing\cite{ref4,ref5} which allow a small mixing
at high scale to be amplified to large mixing at the weak scale after
renormalization group evolution. The only condition that needs to be
fulfilled is that the
$\nu_{\mu}$ and $\nu_{\tau}$ be quasi-degenerate in mass, as for example
would be independently required
if LSND results are confirmed. In such models, there is no need to impose
special constraints on the theory at high scale beyond those needed to
guarantee the quasi-degeneracy. They would, therefore, require less
theoretical input compared to the case where one tries to obtain both
degeneracy (should it be phenomenologically warranted) and maximal mixing
at the high (seesaw) scale within the framework of quark lepton
unification.
\par
A key role in the above scenario is played by the renormalization group
equations for neutrino masses and mixings \cite{ref6,ref7}. In this paper,
we exploit one of the most interesting and highly appealing aspect of
renormalisation group (RG) running of gauge and Yukawa couplings i.e.~the
emergence of fixed point (FP) structure. Use of FP structure of
RGE's to understand low energy parameters is not new; for example, it is
often invoked \cite{ref8} to understand
the top quark mass at the weak scale starting with a wider range of
possible values at the GUT scale as would be preferred by the naturalness
requirement in GUT theories. 
\par
It was noted in \cite{ref6} that neutrino mixings can have fixed points  
corresponding to maximal mixing and several examples were
given to illustrate this point in the standard model and two Higgs model.
The desirable value of  $\sin^2(2\theta)\sim 1$ were shown
to arise in these models both at the electroweak and at intermediate scales 
of order $10^{8}$ GeV or so depending on the model parameters at the 
high scale. Our goal in the present work is to extend the 
discussion of \cite{ref6} to supersymmetric theories (MSSM) and 
delineate the constraints on the high scale theory
under which the fixed point (or maximal mixing) occurs around the weak
scale. We discuss the conditions under which the value of the mixing remains
stable as the energy is varied from TeV to $M_Z$ scale. A
crucial requirement for the fixed point to occur is that the muon and the tau
neutrinos must be quasi-degenerate. Our analysis further clarifies the
idea of radiative magnification discussed in Ref.\cite{ref4,ref5}.
We the point out that in a special class of models which extend this idea
to the case of three
degenerate neutrinos, then searches for neutrinoless double beta decay
can provide a test of these models.
\par
The paper is organised as follows. In section \ref{sec2}, we discuss
radiative
correction and derivation of RGEs for mixing angle in the standard model (SM)
and minimal supersymmetric standard model (MSSM). In section \ref{sec3}, 
we obtain analytic solution for the RGE and demonstrate explicitly the
renormalisation
group fixed point (RGFP) structure. In section \ref{sec4}, we show how the
FP occurs naturally at the weak scale for quasidegenerate neutrinos leading to
the condition of radiative magnification. We also derive a new stability
criterion and show how the FP and magnification occur in MSSM
starting from small mixings as in the quark sector. In section \ref{sec5}, 
we comment on tests of the radiative magnification scheme in
 neutrinoless double beta decay searches.
\par
\section{Radiative Corrections and RGE for Neutrino Mixing}
\label{sec2}
\par
In both SM and MSSM, we consider radiative corrections in
the flavor basis to the light Majorana neutrino mass matrix,
$m_{\alpha\beta}$, which is a 5-dim operator scaled by the high mass, $M_N$
(e.g. see-saw or R.H.~Majorana neutrino mass scale), where the mass matrix is
generated, with 
\be{\cal L}_{\nu\nu}=-\nu_\alpha^TC^{-1}m_{\alpha_\beta}\nu_\beta
+{\rm h.c..}\label{eq1}\ee
\par
As a result of one-loop radiative corrections the RGEs below $\mu=M_N$ are
\cite{ref6,ref7}
\par
\widetext
\noindent\underline{SM}
\ba\dmdt&=&\left[-3g_2^2+2\lambda+{\rm
Tr}\left(6\UU+6\DD+2\EE\right)\right]m\nonumber\\
&-&{1\over 2}\left[m\left(Y_EY_E^\dagger\right)
+\left(Y_EY_E^\dagger\right)^Tm\right]~.\label{eq2}\ea
\par
\noindent\underline{MSSM}
\be\dmdt=\left[-{6\over 5}g_1^2-6g_2^2
+{\rm Tr}\left(6\UU\right)\right]m
+\left[m\left(Y_EY_E^\dagger\right)
+\left(Y_EY_E^\dagger\right)^Tm\right]~.\label{eq3}\ee
\par
\narrowtext
In (\ref{eq2}) and (\ref{eq3}) $g_1(g_2)$ are the $U(1)_Y(SU(2)_L)$ 
gauge couplings, $\lambda$ is the Higgs quartic coupling (in SM), 
whereas $Y_U(Y_D)$ and $Y_E$
are the Yukawa matrices for the up(down) quarks and charged leptons. We work
in the charged lepton diagonal basis where the unitary matrix $U_{\alpha i}$
that transforms mass basis to flavor basis is identified as the standard MNS
matrix \cite{ref9}. Using (\ref{eq2}) and (\ref{eq3}), the mass matrix is 
evolved from the high scale down to $t(=\ln\mu)<t_0(=\ln M_N)$ in the SM or 
MSSM.
\par
\mediumtext
\noindent\underline{SM}
\be m(\mu)=\left({\mu\over M_N}\right)^{\left(\lambda\over 16\pi^2\right)}
I^{-3}_{g_2}I^6_{\rm top}I^3_bI_\tau
\times\left[\br{ccc}
\mee I'_e&\memu\sqrt{I'_e I'_\mu}&\metau\sqrt{I'_eI'_\tau}\\
\mmue\sqrt{I'_eI'_\mu}&\mmumu I'_\mu&\mmutau\sqrt{I'_\mu I'_\tau}\\
\mtaue\sqrt{I'_eI'_\tau}&\mtaumu\sqrt{I'_\mu I'_\tau}&
\mtautau I'_{\tau}\er\right]~.\label{eq4}\ee
\par
\noindent\underline{MSSM}
\be m(\mu)=I_{g_1}^{-6\over 5}I^{-6}_{g_2}I^6_{\rm top}
\times\left[\br{ccc}
\mee I^2_e&\memu I_e I_\mu&\metau I_eI_\tau\\
\mmue I_eI_\mu&\mmumu I^2_\mu&\mmutau I_\mu I_\tau\\
\mtaue I_eI_\tau&\mtaumu I_\mu I_\tau&\mtautau I^2_{\tau}
\er\right]~.\label{eq5}\ee
\narrowtext
Here,
\be I_h(\equiv {I'_h}^{-1})=e^{\delta_h}=\exp\left({1\over 8\pi^2}
\int_{t_0}^{t}h^2(t')dt'\right)~,\label{eq6}\ee
and $h$ denotes the  gauge coupling $(g_1,g_2)$ or the Yukawa-coupling-eigen
value for quarks and charged leptons $(y_{\rm top},y_b,y_{\tau},
y_{\mu},y_e)$. When the running VEV of the up-type Higgs doublet in MSSM is
taken into account the common factor in (\ref{eq5}) is changed with the
replacement, $I_{g_1}^{-6/5}I_{g_2}^{-6} I_{\rm top}^6\to I_{g_1}^{-9/10}
I_{g_2}^{-9/2}$ and similarly in SM. In subsequent discussions for mixing
angle we ignore common renormalisation factors in (\ref{eq4}) and (\ref{eq5}) 
as they cancel out in the relevant expressions. At any value of $t<t_0$,
\be\tan 2\theta(t)={2m_{\mu\tau}(t)\over
m_{\tau\tau}(t)-m_{\mu\mu}(t)}\label{eq7}~,\ee
and (\ref{eq2}), (\ref{eq3}), and (\ref{eq7}) give the RGEs for
$\sin^2(2\theta)$,
\par
\widetext
\noindent\underline{SM}
\ba 16\pi^2{d\sn2t\over dt}&=&\sn2t\cos^2 2\theta\left(y_\tau^2-y_\mu^2\right)
{m_{\tau\tau}+m_{\mu\mu}\over m_{\tau\tau}-m_{\mu\mu}}\nonumber\\
&=&\sn2t{\left(m_{\tau\tau}^2-m_{\mu\mu}^2\right)\left(y^2_\tau-y^2_\mu\right)
\over \left(m_{\tau\tau}-m_{\mu\mu}\right)^2+4m_{\mu\tau}^2}~.\label{eq8}\ea
\par
\noindent\underline{MSSM}
\ba 16\pi^2{d\sn2t\over dt}&=&-2\sn2t\cos^2 2\theta\left(y_\tau^2-y_
\mu^2\right){m_{\tau\tau}+m_{\mu\mu}\over m_{\tau\tau}-m_{\mu\mu}}\nonumber\\
&=&-2\sn2t{\left(m_{\tau\tau}^2-m_{\mu\mu}^2\right)\left
(y^2_\tau-y^2_\mu\right)
\over \left(m_{\tau\tau}-m_{\mu\mu}\right)^2+4m_{\mu\tau}^2}~.\label{eq9}\ea
\narrowtext
All quantities in the RHS of (\ref{eq8}) and (\ref{eq9}) are $t$-dependent. 
As was noted in \cite{ref6}, both the RGEs have one trivial fixed point at
$\sin^22\theta=0$ and the other nontrivial fixed point at $\sin^22\theta=1$.
Recently the FP structure of MNS matrix has been also investigated in
\cite{ref10}. Assuming that the initial high-scale texture of the mass matrix
is such that the nontrivial fixed point occurs at a scale
$\mu_c (M_Z\leq\mu_c < M_N, t_c=\ln\mu_c)$, we have the FP condition,
\be\sn2t(t_c)=1~,\label{eq10}\ee
or, equivalently,
\be m_{\tau\tau}(t_c)=m_{\mu\mu}(t_c)~.\label{eq11}\ee
\par
\section{Analytic Formula and Fixed Point}
\label{sec3}
\par
Before obtaining analytic solutions to (\ref{eq8}) and (\ref{eq9}), it is
worthwhile to explain why resonance structures in the numerical solutions
\cite{ref6,ref10,ref11,ref12} in the $\sn2t(t)$ vs.~$t$ plots are expected for
specific textures of $m^0_{\alpha\beta}$.

 Noting that,
\be\left|y_\tau(t)\right|^2\gg\left|y_\mu(t)\right|^2\label{eq12}\ee
(\ref{eq5}) states that in MSSM, as $t$ decreases below $t_0$, the ratio,
$R_\tau(t)=m_{\tau\tau}(t)/m_{\tau\tau}^0$, decreases faster from its high
scale value, $(R_\tau(t_0)=1)$, as compared to the rate of decrease of the
ratio, $R_\mu(t)=m_{\mu\mu}(t)/m_{\mu\mu}^0 $. In particular, the relations
(\ref{eq10}) and (\ref{eq11}) are satisfied at $t=t_c$ if
\be m_{\tau\tau}^0e^{2\delta_\tau(t_c)}=m_{\mu\mu}^0~.\label{eq13}\ee
For the FP to occur at $t_c<t_0$, the high scale texture must be such that
$m_{\tau\tau}^0$ and $m_{\mu\mu}^0$ are comparable but unequal with
$m_{\tau\tau}^0>m_{\mu\mu}^0$. Lower values of $t_c$ corresponds to larger
differences between $m_{\tau\tau}^0$ and $m_{\mu\mu}^0$.
\par
From (\ref{eq9}) it is clear that when $m_{\tau\tau}(t_c)=m_{\mu\mu}(t_c)$,
the slope of the curve in the $\sn2t(t)$ vs.~$t$ plot vanishes at $t=t_c$.
For $t>t_c$, $m_{\tau\tau}(t)>m_{\mu\mu}(t)$, the slope is negative; but for
$t<t_c$, $m_{\tau\tau}(t)<m_{\mu\mu}(t)$ and the slope is positive as given
by the RGE. Negative(positive) slope to the right(left) with vanishing slope
at $t=t_c$ is the characteristic feature of a resonance curve as predicted
by (\ref{eq9}) for MSSM. Similar result emerges for SM from (\ref{eq8}) with
somewhat different high-scale condition with $\mmumu>\mtautau$ and the ratios
$R_\tau(t)$ and $R_\mu(t)$ increase as $t$ decreases below $t_0$. Thus it is
clear that for certain given textures at high scale $(\mmumu, \mtautau$ and
$\mmutau)$ resonance occurs at $t=t_c$ converting small mixing at high scale
to large mixing at lower scales.
\par
Inspite of the terse nature of the RHS of (\ref{eq9}), using the almost
exact approximation, $|\delta_\tau(t)|\gg|\delta_\mu(t)|$, we have
integrated
it to obtain analytic solution for RG evolution of $\sn2t$ in the MSSM for
all values of $\mu<M_N$,
\mediumtext
\be\sn2t(t)=\sn2t_0{\left[\left(\mtautau-\mmumu\right)^2
+4{\mmutau}^2\right]e^{2\delta_\tau(t)}
\over \left[\left(\mtautau e^{2\delta_\tau(t)}
-\mmumu\right)^2+4{\mmutau}^2e^{2\delta_\tau(t)}\right]}~,\label{eq14}\ee
\narrowtext
where, $\theta_0$, is the high scale mixing angle with
\be\tan 2\theta_0={2\mmutau\over \left(\mtautau-\mmumu\right)}~.\label{eq15}\ee
\\
Given,
\ba m_{\mu\mu}(t)/m_{\tau\tau}(t)&=&(\mmumu/\mtautau)
e^{2\delta_\mu (t)}e^{-2\delta_
\tau (t)}~,\nonumber\\
m_{\mu\tau}(t)/m_{\tau\tau}(t)&=&(\mmutau/\mtautau)
e^{\delta_\mu (t)}e^{-\delta_
\tau (t)}~,\nonumber\ea
and using (\ref{eq5})-(\ref{eq7}), (\ref{eq14}) may be recognised
as the approximation, $|\delta_\mu(t)|\ll|\delta_\tau(t)|$, to the following
exact analytic solution of (\ref{eq9}),
\mediumtext
\be\sn2t(t)=\sn2t_0{\left[\left(\mtautau-\mmumu\right)^2+4{\mmutau}^2\right]
e^{2\delta_\tau(t)}e^{2\delta_\mu(t)}\over
\left[\left(\mtautaut-\mmumut\right)^2+4\mmutaut^2\right]}~.\label{eq16}\ee
\narrowtext
Replacing $\theta(t)\to\theta(\mu)$, $m_{ij}(t)\to m_{ij}(\mu)$, $\theta_0
\to\theta(M)$, and $m_{ij}^0\to m_{ij}(M)$, formulas (\ref{eq14}) or
(\ref{eq16}) can be used to derive $\theta(\mu)$ from $\theta(M)$ or vice
versa for all values of $\mu < M \le M_N$. 
It is interesting to note that these analytic solutions exhibit both the
resonance as well as  the nontrivial FP structure explicitly. While detailed
features of resonance such as the $t$-dependent width, maximal mixing at the
peak, and smaller mixings for $t>t_c$ or $t<t_c$ are clearly exhibited, the
FP structure is proved as follows. At $t=t_c$, when (\ref{eq11}) or 
(\ref{eq13}) are satisfied, the quantity inside the parenthesis in the 
denominator of (\ref{eq16}) or (\ref{eq14}) vanishes. Then using
(\ref{eq15}), (\ref{eq14}) and (\ref{eq16}) give
\mediumtext
\be\sn2t(t_c)=\sn2t_0\left[{\left(\mtautau-\mmumu\right)^2\over
{4\mmutau}^2}+1\right]=\sn2t_0+\cos^22\theta_0=1~.\label{eq17}\ee
\narrowtext
It is to be noted that (\ref{eq17}) holds for all initial value of
$\theta_0<\pi/4$, thus demonstrating the fixed point behavior corresponding
to maximal mixing. Although the relation (\ref{eq17}) appears to be true also
for $\theta_0=\pi/4$ showing that maximal mixing remains maximal at $t=t_c$,
the RG evolution equations never satisfy $m_{\tau\tau}(t_c)=m_{\mu\mu}(t_c)$
for $t_c\ll t_0$ if we start with the initial condition $m_{\tau\tau}^0=
m_{\mu\mu}^0$ which is necessary for $\theta_0=\pi/4$. In fact nearly maximal
mixings at the high scale are damped out to small mixings at lower scales
$(\mu\ll M_N)$ due to nonvanishing contributions of the quantity
$\left(m_{\tau\tau}(t_c)-m_{\mu\mu}(t_c)\right)^2$ in the R.H.S.~of
(\ref{eq14}) or (\ref{eq16}). Thus the analytic formula, apart from
demonstrating the FP structure and resonance behavior, also explains why
large mixing at high scales are damped out to small mixings near the weak
scale. Also zero mixing angle  does not run and continues to be zero down to
$\mu=M_Z$. Similar analytic solutions are also obtained for SM exhibiting the
FP structure with the replacement $2\delta_i(t) \to - \delta_i(t)$,
$i=\mu,\tau$ in (\ref{eq14})-(\ref{eq16}).
\par
In almost all cases of RG fixed point discussed so far in the literature, the
FP structure is revealed through the differential RGEs and demonstrated
through numerical solutions only. But in the present case, apart from the
differential RGE and numerical solutions (see section \ref{sec4}), the
analytic solutions also exhibit the FP structure explicitly as demonstrated
through (\ref{eq14})-(\ref{eq17}).
\par
\section{Radiative Magnification Through the Fixed Point and Stability}
\label{sec4}
\par
When the condition (\ref{eq10}) or (\ref{eq11}) is satisfied for $t_c=t_S=\ln
M_S$($M_S=$SUSY scale), the FP may manifest as large neutrino mixing observed
at low energies, for example, in $\nu_\mu-\nu_\tau$ oscillation scenario
necessary to solve the atmospheric neutrino anomaly. In terms of the high
scale mass eigen values $(m_2^0,m_3^0)$, mixing angle $(\theta_0)$, and
radiative correction parameters, the condition for FP manifestation at
$\mu=\mu_c=M_S$ then reduces to
\mediumtext
\be (m_2^0-m_3^0)c_{2\theta_0}=2\delta_\tau(t_S)\left(m_2^0s^2_{\theta_0}+
m_3^0c^2_{\theta_0}\right)-2\delta_\mu(t_S)\left(m_2^0c^2_{\theta_0}+
m_3^0s^2_{\theta_0}\right)~.\label{eq18}\ee
\narrowtext
where $s_{\theta_0}=\sin\theta_0$, $c_{\theta_0}=\cos\theta_0$,
$c_{2\theta_0}=\cos 2\theta_0$ and $s_{2\theta_0}=\sin 2\theta_0$. Taking
$M_S=M_Z$, this is recognised 
exactly as the condition that was derived in \cite{ref4,ref5} for magnifying
small mixing at high scale to large mixing at low-energies through radiative
corrections. But, as noted here, the condition is exact, needs no fine tuning,
and emerges as a natural consequence of the manifestation of the FP at the
weak scale. For small mixing angles at $\mu=M_N$, similar to those existing in
the quark sector (e.g.~$\theta_0\approx V_{cb}\approx 0.04)$,
$c_{\theta_0}\approx c_{2\theta_0}\approx 1$, and $s^2_{\theta_0}\sim0$, it is
clear that the condition (\ref{eq18}) cannot be satisfied if the masses
$m_2^0$ and $m_3^0$ are hierarchial, or exactly degenerate having the same
$(m_2^0=m_3^0)$ or opposite CP-parity $(m_2^0=-m_3^0)$. Also it cannot be
satisfied if the masses are quasidegenerate with opposite CP-parity
$(m_2^0\sim -m_3^0)$. It can be satisfied only if the masses are
quasidegenerate at the high scale having the same CP-parity
$(m_2^0\sim m_3^0)$. Since $\delta_\tau$ is negative, a necessary prediction
of MSSM is that $m_3^0>m_2^0$. In the SM, $2\delta_\tau(t_Z)$ and
$2\delta_\mu(t_Z)$ in (\ref{eq18}) are replaced by $-\delta_\tau(t_Z)$ and
$-\delta_\mu(t_Z)$, respectively, and (\ref{eq18}) predicts $m_2^0>m_3^0$.
These requirements in MSSM or SM are analogous to the occurrence of quasi
fixed points in top-quark Yukawa coupling where right order of the top quark
mass is obtained only for certain strong interaction couplings. We emphasize
that the observed large neutrino mixing in the $\nu_\mu-\nu_\tau$ sector
predicts the corresponding $\nu_2(\nu_3)$ masses to be quasidegenerate with
the same CP-parity as a necessary requirement in order that the FP manifests
at the lower scale. Under the condition (\ref{eq11}), with $t_c=t_S$,
the mass eigen values at $\mu=M_S$ are
\mediumtext
\ba m_2(t_S)&=&\left(m_2^0c_{\theta_0}^2+m^0_3s_{\theta_0}^2\right)
(1+2\delta_\mu(t_S))-(m_3^0-m_2^0)c_{\theta_0}s_{\theta_0}
(1+\delta_\tau(t_S)+\delta_\mu(t_S))~.\label{eq19}\\
m_3(t_S)&=&\left(m_2^0c_{\theta_0}^2+m^0_3s_{\theta_0}^2\right)
(1+2\delta_\mu(t_S))+(m_3^0-m_2^0)c_{\theta_0}s_{\theta_0}
(1+\delta_\tau(t_S)+\delta_\mu(t_S))~.\label{eq20}\ea
\narrowtext
Taking the high scale mixings to be small, we obtain the
mass squared difference at $\mu=M_S$,
\be\Delta m^2\equiv m_3^2-m_2^2\approx\Delta{m^0}^2s_{2\theta_0}
(1+\delta_\tau(t_S))~,\label{eq21}\ee
where,
\be\Delta{m^0}^2\approx 2{m_2^0}^2\left(e^{-2\delta_
\tau(t_S)}-1\right)\approx -4{m_2^0}^2\delta_\tau(t_S)~.\label{eq22}\ee
\par
\begin{figure}
\epsfxsize=16cm
{\epsfbox{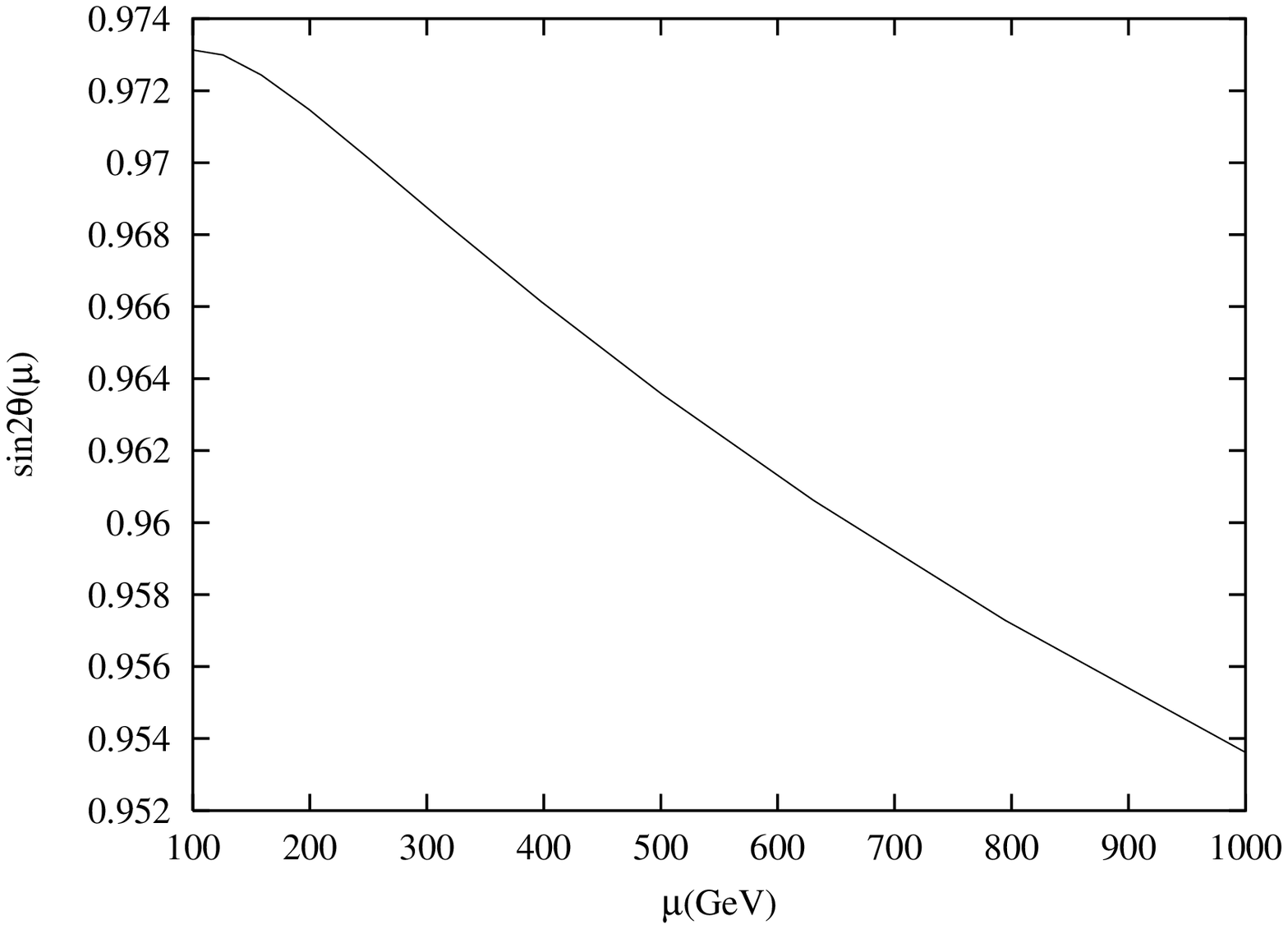}}
\caption{Manifestation of the fixed point at the weak scale and stability of
radiative magnification of small high scale mixing $\sin\theta_0\approx 0.22$
to nearly maximal mixing at low energies for $y_\tau\approx 0.48$,
$\tan\beta\approx 50$, $\mmumu\approx m_2^0\approx 0.26$ eV and
$\mtautau\approx m_3^0\approx 0.28$ eV, and $\mmutau=0.0044$ eV consistent
with atmospheric neutrino data ($\Delta m^2\approx 4\times 10^{-3}$ eV$^2$).}
\label{fig1}
\end{figure}
\begin{figure}
\epsfxsize=16cm
{\epsfbox{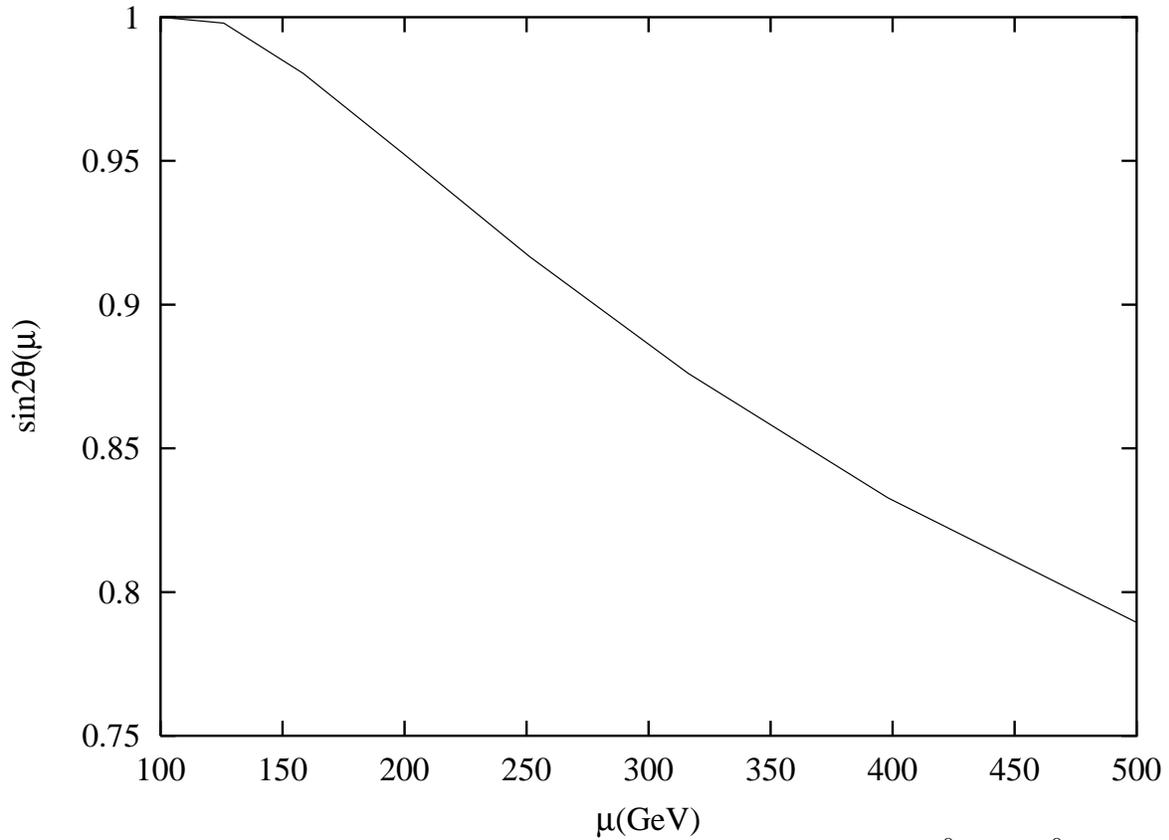}}
\caption{Same as Fig.~\ref{fig1} but for $\sin\theta_0\approx V_{cb}\approx
0.04$, $y_\tau\approx 0.49$, $\mmumu\approx m_2^0\approx 0.25$ eV,
$\mtautau\approx m_3^0\approx 0.27$ eV, $\mmutau\approx 0.0008$ eV and
$\Delta m^2\approx 7\times 10^{-4}$ eV$^2$ consistent with atmospheric
neutrino data.}
\label{fig2}
\end{figure}

Before proceeding further, we show  analytically how the stability of
radiative magnification is controlled by the high-scale mixing angle. To
generate nearly maximal mixing at a lower scale $(\mu=\mu_c=M_S=M_Z)$ starting
from small mixing as in the quark sector at the high scale
(e.g.~$\theta_0\approx V_{cb}\approx 0.04$), the FP position is desired to be
stable near $M_S=M_Z$. As the FP is a consequence of radiative corrections,
the stability must be guaranteed against smaller changes in the neutrino mass
matrix due to higher order corrections. To maintain such stability this
requires the mixing to be nearly maximal within at least $\mu\sim$ few
$(M_Z)$. In fact we show that radiative stability is ensured over a larger
range. We define the range, $t=t_S$ to $t_\Gamma$ $(\mu=M_S$ to $\mu_\Gamma)$,
within which, the mixing remains nearly maximal. Noting that,
\be\delta_\tau(t_\Gamma)=\delta_\tau(t_S)+\epsilon_\tau(t_\Gamma)~,
\label{eq23}\ee
with
\be\epsilon_\tau(t_\Gamma)\approx{y_\tau^2\over 16\pi^2}\ln{\mu_\Gamma\over
M_S}~,\label{eq24}\ee
which remains small $(|\epsilon_\tau |\ll1)$ over a wide range of $y_\tau$,
we use the FP condition (\ref{eq11}) and (\ref{eq13}) in (\ref{eq14})
and (\ref{eq16}) to obtain
\be\sn2t(t_\Gamma)\approx{1+2\delta_\tau(t_\Gamma)\over\left[1+2\delta_\tau
(t_\Gamma)+{{\mmumu}^2\over{\mmutau}^2}\epsilon^2(t_\Gamma)\right]}~.
\label{eq25}\ee
The stability criterion for the FP position and radiative magnification at
$\mu\approx M_S$ may be stated as
\be{y_\tau^4{\mmumu}^2\over 256\pi^4{\mmutau}^2}\left(\ln{\mu_\Gamma\over
M_S}\right)^2\ll 1~.\label{eq26}\ee
This clearly has the implication that arbitrarily small values of high-scale
mixing cannot maintain a stable FP whereas zero initial mixing continues to
remain zero at all lower values of $t$ and is never magnified. For smaller
values of $\theta_0$ or $\mmutau$, the contribution of the third term in the
denominator in (\ref{eq25}) becomes larger leading to sharper decrease of the
predicted low-scale mixing angle from its maximal fixed point value. This
results in the smaller width of the resonance for smaller values of high-scale
mixing ($\theta_0$ or $\mmutau$). This feature is clearly exhibited through
Figs.~\ref{fig1} and \ref{fig2}, where
we have presented $\sin 2 \theta(\mu)$  for  $\mu=100$ GeV-$1$ TeV taking
$M_S=M_Z$, $M_N=10^{13}$ GeV, $\tan\beta=50$ and $y_{\tau}=0.49$ with
$e^{2\delta_{\tau}(M_Z)}=0.929$. The high-scale parameters for Fig.~\ref{fig1}
are $\mtautau\approx m_3^0\approx 0.28$ eV,
$\mmumu\approx m_2^0\approx 0.26$ eV, and $\mmutau \approx 0.0044$ eV
corresponding to $\theta_0=0.22$ consistent with
$\Delta m^2\approx 4\times 10^{-3}$ eV$^2$ needed for atmospheric neutrino
data. For Fig.~\ref{fig2} these parameters are
$\mtautau \approx m_3^0 \approx 0.27$ eV, $\mmumu \approx m_2^0
\approx 0.25$ eV, and $\mmutau \approx 0.0008$ eV corresponding to
$\theta_0=V_{cb}=0.04$ consistent with $\Delta m^2 \approx 7 \times 10^{-3}$
eV$^2$. It is clear that in Fig.~\ref{fig2} the width is substantially
narrower than Fig.~\ref{fig1} and $\sin2\theta(\mu)$ reduces by nearly $20\%$
from its maximal value over the range of $100-500$ GeV. Such energy dependent
mixing between the two neutrinos, as a prediction of MSSM when both the FP and
the SUSY scale are at $M_S=M_Z$ might be possible to testify or falsify in
future by high-energy neutrino experiments.
\par
In contrast to the energy dependent mixing discussed above, for the first time
we find here a very attractive new feature of the other class of MSSM with
higher SUSY scale $M_S=O$(TeV) where stable and almost energy independent
mixing, close to its maximal value, is predicted over a wider range of
energy scale $\mu=M_Z$-few TeV starting from the high-scale mixing similar to
the quark sector, $\theta_0=V_{cb} \approx 0.04$. Using the technique
explained above, the high-scale parameters are chosen to have the RG fixed
point at the SUSY scale $M_S=1$ or few TeV. Then the origin of negligible
energy dependence in the predicted mixing at all lower energy scales is
explained by noting the nonSUSY SM prediction for which
$y_{\tau}\approx 0.01$ below $M_S$,
\par
\begin{figure}
\epsfxsize=16cm
{\epsfbox{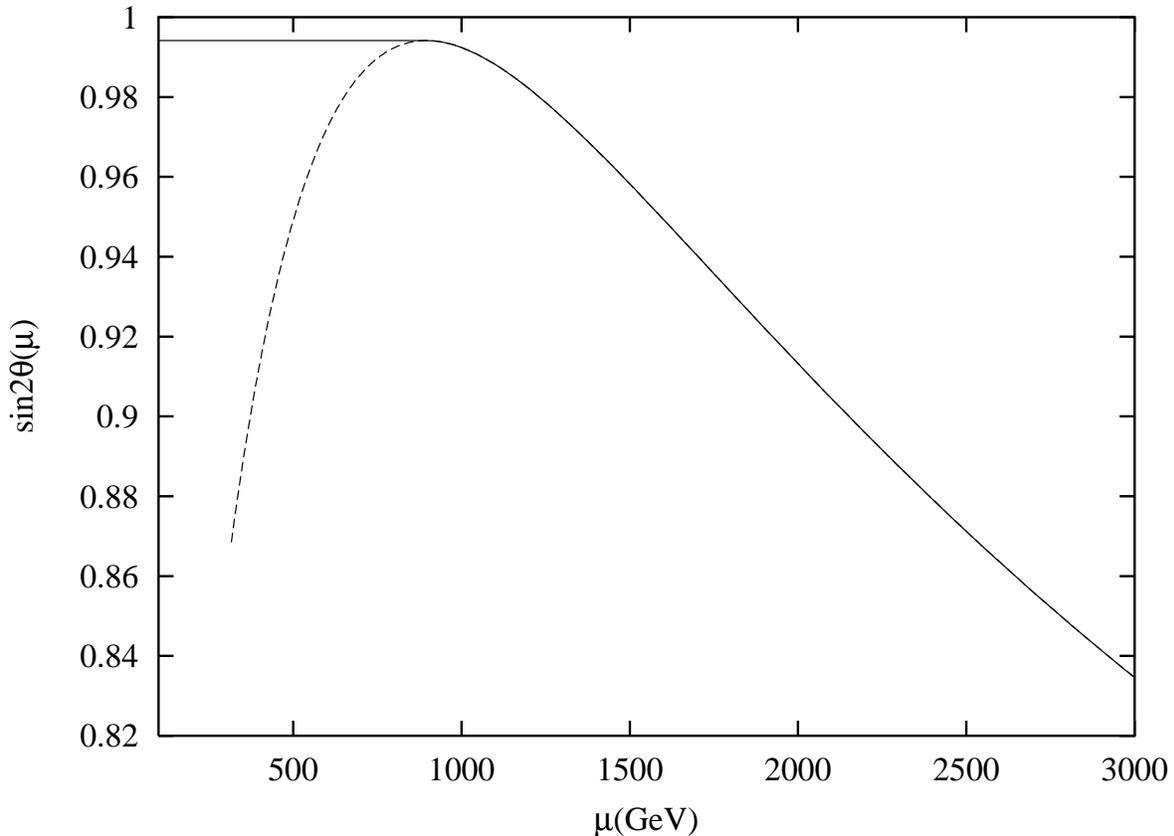}}
\caption{Radiative magnification and an explicit demonstration of stability 
of large neutrino mixing at lower scales for $\mu\le M_S$. The high-scale 
parameters are $M_N=10^{13}$ GeV, $\sin\theta_0=V_{cb}=0.04$, $\mtautau\approx
m_3^0=0.1543$ eV, $\mmumu\approx m_2^0=0.1434$ eV, $\mmutau\approx 0.00044$ eV
with $\Delta m^2\approx 4\times 10^{-3}$ eV$^2$. The dashed line is the
continuation of the resonance curve with the FP at $\approx 1$ TeV and SUSY
scale $M_S=M_Z$. The part of solid line almost flat below 1 TeV has been
obtained with both the FP and SUSY scale at $\approx 1$ TeV. The value of
$y_{\tau}$ is $0.49(0.01)$ for MSSM(SM) corresponding to $\tan\beta=50$.}
\label{fig3}
\end{figure}

\noindent\underline{SM:} $M_Z\le \mu \le M_S$,
\mediumtext
\ba \sn2t(\mu)&=&\sn2t(M_S){\left[\left(m_{\tau\tau}(M_S)
-m_{\mu\mu}(M_S)\right)^2+4m_{\mu\tau}^2(M_S)\right]
\left({M_S\over\mu}\right)^{y_\tau^2\over 16\pi^2}\over\left(m_{\tau\tau}(M_S)
\left({M_S\over\mu}\right)^{y_\tau^2\over 16\pi^2}-m_{\mu\mu}(M_S)\right)^2+
4m_{\mu\tau}^2(M_S)\left({M_S\over\mu}\right)^{y_\tau^2\over 16\pi^2}}
\label{eq27}\\
&=&\sn2t(M_S){4m_{\mu\tau}^2(M_S)
\left({M_S\over\mu}\right)^{y_\tau^2\over 16\pi^2}\over m_{\mu\mu}(M_S)\left(
\left({M_S\over\mu}\right)^{y_\tau^2\over 16\pi^2}-1\right)^2+
4m_{\mu\tau}^2(M_S)\left({M_S\over\mu}\right)^{y_\tau^2\over 16\pi^2}}~,
\label{eq28}\ea
\narrowtext
where, (\ref{eq28}) has been obtained from (\ref{eq27}) by using the
FP condition, $m_{\tau\tau}(M_S)=m_{\mu\mu}(M_S)$. Then, because of smallness
of the $\tau$-Yukawa coupling in the SM with $y_{\tau} \simeq 0.01$ in
(\ref{eq28})
there is negligible $\mu$-dependence and the predicted mixing remains stable,
close to its maximal value, for all values of $\mu$ below $M_S=O$(TeV). This
behavior is shown in Fig.~\ref{fig3} for initial values
of $\theta_0=V_{cb}=0.04$, $m_3^0 \approx \mtautau=0.1543$ eV,
$m_2^0\approx \mmumu=0.1434$ eV, $\mmutau=0.00044$ eV, $\Delta m^2=4\times
10^{-3}$ eV$^2$ and other values of parameters same as in Figs.~\ref{fig1} and
\ref{fig2}, but now having the FP at $M_S \approx 1$ TeV. The dashed line of
Fig.~\ref{fig3} shows the continuation of the resonance structure at
$\mu_c\approx 1$ TeV when the SUSY scale is $M_Z$ and the evolution of mixing
throughout is as in MSSM given by (\ref{eq14}) or (\ref{eq16}). The part of
the solid line below $\mu\approx 1$ TeV exhibiting almost flat
behavior of the predicted mixing angle, with $\sin2\theta(\mu)\approx 0.99$,
has been obtained using
(\ref{eq28}) with $y_\tau=0.01$ and corresponds to the FP and the SUSY scale
both at $M_S\approx 1$ TeV. In this case the formula (\ref{eq25}) applies to
the part of the curve above $M_S\approx 1$ TeV. Thus, we have shown for the
first time that after radiative magnification through manifestation of fixed
point at $\mu=M_S=O$(TeV), the predicted mixing remains stable and close
to its maximal value at all lower energy scales. In this regard our analysis
favors the class of MSSM with $O$(TeV) SUSY scale.
\par
As explained in \cite{ref4,ref5} while keeping the quasidegenerate eigen
states $\nu_2$ and $\nu_3$ to have the same CP-parity for radiative
magnification, it is necessary to have CP-parity of $\nu_1$ to be opposite to
prevent radiative magnification in the $\nu_e-\nu_\tau$ sector from small
values of $\theta_{13}$ which are consistent with CHOOZ-PALOVERDE 
\cite{ref16} bound. In this case solar neutrino anomaly \cite{ref13,ref14} 
is explained by $\nu_e\to\nu_\mu$ oscillation through small angle MSW effect. 
\par
\section{Testing radiative magnification by neutrinoless double beta
decay}
\label{sec5}
\par
 In this section, we briefly remark on the implications of our
magnification scheme for neutrinoless double beta decay experiments. 

So far we have considered only two
generation mixing. In complete models, one will have to embed this
mechanism into scenarios
with  three generations or three generations plus a sterile neutrino. In
the former case, if $\nu_{\mu}$ and $\nu_{\tau}$ are degenerate, then we
have all three neutrinos nearly degenerate in mass in order to fit solar
and atmospheric neutrino data. In particular, we could have all three
neutrinos to have same CP. An example of such an extension is given in
\cite{ref5}. We see below that in this particular embedding of our
scenario, neutrinoless double beta decay can provide a test of the idea
of radiative magnification of the atmospheric neutrino mixing.

Neutrinoless double beta decay experiment measures
\be
m_{ee} = \sum_k U_{ek}^2 m_k~,
\label{eq29}
\ee
where, $k$ denotes the mass eigenstate label. For our case with $U_{ek}\ll
1$, $m_{ee}\simeq m_0$ where $m_0$ is the common mass of all the
neutrinos. We will show now that for radiative magnification scheme to
work, one must have a lower limit on the common mass of all neutrinos 
$m_0$ which depends on the value of
$\tan\beta$ of the MSSM. From (\ref{eq22}), we have the lower
bound 
\be
\frac{4 \pi^2 v^2 \Delta m_{atm}^2}{m_{\tau}^2(1+\tan^2\beta)\ln(M_N/M_S)}
\leq m_0^2
\label{eq30}
\ee
In Fig. 4, for a fixed $M_N/M_S$, we show the variation of the lower 
bound on $m_0$ with $\tan\beta$. We have choosen $M_N = 10^{13}$ GeV, 
$M_S = 1$ TeV and we see that for 
lower $\tan\beta$ values, the lower bound on the common mass increases.
Infact, for small initial mixings, a lower common mass implies a 
larger $\tan\beta$. For large $\tan\beta \approx 50-60$, with $M_N
= 10^{13}$ GeV and $M_S= 1$ TeV, the lower bound on the common mass varies
in the range, $m_0 \approx 0.18-0.20$ eV. Thus,  once
supersymmetry is discovered and the value of $\tan\beta$ is
determined,
combining this with the improved searches for neutrinoless double beta
decay\cite{genius}, one can test the idea of radiative magnification for 
the three generation model. In particular, note that the lower limit on
$m_0$ predicted above is very near the present upper limits. This should
provide strong motivation to improve the limits on the lifetime of
neutrinoless double beta decay.
 
\par
\section{Conclusion}
\label{sec6}
\par
We presented the analytic formula for RG evolution of neutrino mixing which
demonstrates explicitly the FP structure corresponding to maximal neutrino
mixing at the weak scale leading to the condition of radiative magnification.
We have derived stability criterion for radiative magnification and show 
that the radiatively magnified two-neutrino mixing, predicted by the RG 
fixed point structure, remains stable and close to its maximal value, for 
all  energy scales below the $O$(TeV) SUSY scale in MSSM. This result is 
specific to the MSSM and cannot be realized in nonSUSY SM. When this 
mechanism is applied to the
standard model, one gets only a resonance structure with maximal mixing at
$M_Z$ and smaller mixings at all higher scales which are energy dependent.
Our numerical computations with the help of the analytic
formulae clearly show that radiative magnification of high scale neutrino
 mixing takes place for quasidegenerate neutrinos having the same
CP-parity and it remains stable only for the MSSM. We point out a very
interesting test of the three generation embedding of this model 
by improving limits on the common mass $m_0$
from $0\nu \beta \beta$ searches \footnote{We thank A.Yu.~Smirnov for
pointing out this feature to us.}. 

\par
\acknowledgments
\par
The work of R.N.M is supported by the NSF grant No.~PHY-9802551. The work of
M.K.P is supported by the DAE project No.~98/37/9/BRNS-Cell/731 of the
Govt.~of India. M.K.P also thanks Y.~Achiman for useful discussions. We thank 
Goran Senjanovi\'c for a critical reading of the manuscript and 
Alexei Yu Smirnov for many clarifications on related issues. Balaji wishes
to thank the high energy theory group at The Abdus Salam ICTP for local 
support and Abdel Perez Lorenzana for many enjoyable discussions.
\par
\narrowtext

\par
\narrowtext
\par
\narrowtext
\par
\narrowtext
\par
\narrowtext
\begin{figure}
\epsfxsize=16cm
{\epsfbox{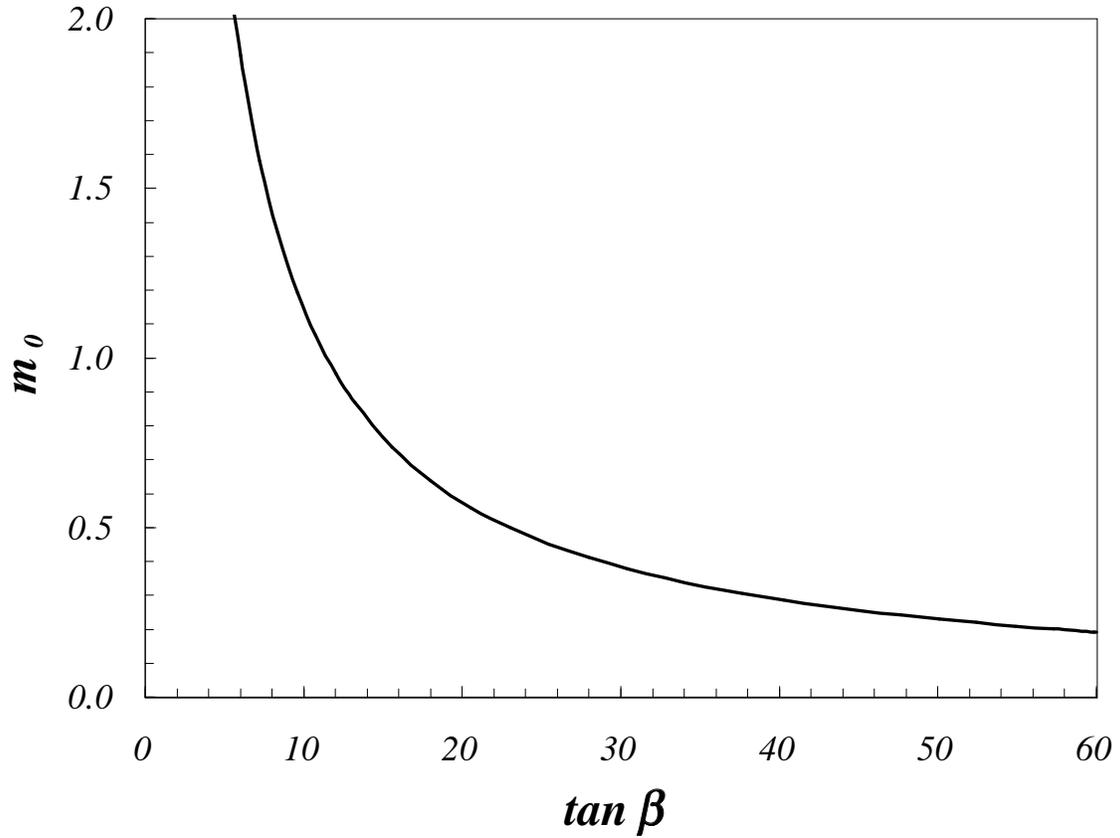}}
\caption{Allowed lower bounds for the common mass, $m_o$ (in eV), for varying 
$\tan\beta$, with $\Delta m_{atm}^2 \approx 4\times 10^{-3}$ eV$^2$,~ 
$M_N = 10^{13}$ GeV 
and $M_S=1$ TeV. For any other value of $M_N/M_S$, the corresponding 
lower value for $m_o$ scales accordingly.}
\label{fig4}
\end{figure}

\end{document}